\documentclass[12pt,a4paper]{article}
\usepackage{epsfig,amssymb}

\newcommand{\be}{\begin{equation}}
\newcommand{\ee}{\end{equation}}
\newcommand{\bea}{\begin{eqnarray}}
\newcommand{\eea}{\end{eqnarray}}

\newcommand{\pa}{\partial}

\newcommand{\R}{{\bf R}}
\newcommand{\Z}{{\bf Z}}

\newcommand{\bJ}{\bar{J}}

\newcommand{\cC}{{\cal C}}

\newcommand{\cZ}{{\cal Z}}

\newcommand{\mg}{{\mathfrak g}}
\newcommand{\mh}{{\mathfrak h}}
\newcommand{\rU}{{\rm U}}
\newcommand{\rO}{{\rm O}}
\newcommand{\rSU}{{\rm SU}}
\newcommand{\rSO}{{\rm SO}}
\newcommand{\rSp}{{\rm Sp}}
\newcommand{\msu}{\mathfrak{su}}

\newcommand{\dfrac}[2]{\displaystyle\frac{\mathstrut #1}{\mathstrut #2}}

\begin{document}

\thispagestyle{empty}
\addtocounter{page}{-1}

\begin{flushright}
{\tt hep-th/0207077}\\
\end{flushright}

\vskip 2cm

\begin{center}
\Large \bf Classification and Quantum Moduli Space of D-branes
in Group Manifolds
\end{center}

\vskip 1cm

\begin{center}
\large Taichi Itoh ~and~ Sang-Jin Sin
\end{center}

\begin{center}
\it Department of Physics, Hanyang University, Seoul 133-791, Korea
\end{center}

\vskip 1cm

\begin{abstract}
We study the classification of D-branes in all compact Lie groups
including non-simply-laced ones. We also discuss the global
structure of the quantum moduli space of the D-branes. D-branes
are classified according to their positions in the maximal torus.
We describe rank 2 cases, namely $B_2$, $C_2$, $G_2$, explicitly
and construct all the D-branes in $B_r$, $C_r$, $F_4$ by the
method of iterative deletion in the Dynkin diagram. The discussion
of moduli space involves global issues that can be treated in
terms of the exact homotopy sequence and various lattices. We also
show that singular D-branes can exist at quantum mechanical level.
\end{abstract}

\vspace*{\fill}
\hrule

\vskip 0.2cm
\noindent
taichi@hepth.hanyang.ac.kr,\\
sjs@hepth.hanyang.ac.kr

\baselineskip=18pt
\newpage


\noindent
{\bf 1. Introduction}
\vspace*{3pt}

Group manifolds provide us solvable string theory backgrounds
in terms of current algebra. Through gluing the left and right chiral
currents up to automorphisms, the (twisted) conjugacy classes therein
turn out to be D-branes \cite{AS,FFFS,S}.\footnote{
Not all the D-branes in group manifolds are given like this.
For example, see section 4 in \cite{FFFS}.}
Though there is an extensive literature on this subject, it is mostly
on either the generic brane with the highest possible dimension or
D0-brane with the lowest one. However, recognizing D-branes as the
conjugacy classes, we notice that there is a variety of D-branes between
these two extremes. In a recent paper, Stanciu \cite{S3} studied the
singular D-branes in SU(3) case.
We developed in \cite{IS} the classification and the systematic construction
of all possible untwisted D-branes in Lie groups of A-D-E series.
D-branes are classified according to their positions
in a unit cell of the weight space which is exponentiated
to be the maximal torus. However, for the D-brane classification,
we only have to consider the fundamental Weyl domain that is surrounded by
the hyperplanes defined by Weyl reflections. All the D-branes therein
can be constructed by the method of iterative deletion in the Dynkin diagram.
The dimension of a D-brane always becomes an even number and
it reduces as we go from a generic point of the fundamental domain to its
higher co-dimensional boundaries.

In this paper, we first generalize the classification of D-branes for
non-simply-laced compact Lie groups and then discuss the global structure
of the quantum moduli space of the D-branes.
Rank 2 cases, namely $B_2$, $C_2$, $G_2$, are discussed explicitly and
all the D-branes for $B_r$-, $C_r$-series and $F_4$ are constructed
by the Dynkin diagram method.
For non-simply-laced cases, the periods of central lattice are different
for short and long roots so that not all the vertices of the fundamental
domain correspond to the D0-branes, resulting in a richer variety of
D-brane Zoo compared with simply-laced cases.
For $B_r$-series, the discussion of D-brane moduli space involves the
global structure of the groups that comes from the difference between
the integral lattice and the co-root lattice. For example, SO(5) and Sp(2)
share the same fundamental domain but they are different in the period
of integral lattice, resulting in different topological structures.

\vspace*{12pt}
\noindent
{\bf 2. Classifying the D-branes}
\vspace*{3pt}

The chiral $G_L \times G_R$ symmetry of WZW model is generated by
the left and right chiral currents, $J = -\pa_+ g g^{-1}$,
$\bJ = g^{-1} \pa_- g$, which induce translations on the group manifold $G$
by the left-right action $g \mapsto l g\hspace{0.5pt} r^{-1}$
with $(l,r)\in (G_L,G_R)$. We are interested in the world sheet boundary
conditions preserving half the chiral $G_L \times G_R$ symmetry
\cite{ishibashi,KO}. Such a boundary condition may be given by identifying
$J$ with $\bJ$ up to automorphisms $\Omega$ of the Lie algebra
$\mathfrak g$ \cite{AS,FFFS,S}: $J =\Omega(\bJ)$ at $\sigma=0$.
This gluing condition restricts the left-right action to a (twisted)
conjugation: $g \mapsto \omega(r) g\hspace{0.5pt} r^{-1}$ with $r \in G_R$,
where $\omega$ is generated by $\Omega$. As $r$ runs over the entire group,
the conjugation action translates open string end points to all over the
conjugacy class. Therefore we may identify a D-brane or the set of open
string end points as the conjugacy class \cite{AS,FFFS,S}.

Since any conjugacy class pass through a point in the maximal
torus $T$, (or its invariant subgroup
$T^\omega =\{t \in T|\,\omega(t)=t\}$ if $\omega\ne 1$)
we can parameterize the
conjugacy classes by the elements $h$'s in the maximal torus:
\be
\cC_\omega (h)=\{\omega(g)hg^{-1}~\mbox{with}~g \in G\}.
\label{conj}
\ee
The dimension of a D-brane thus depends on the symmetry group of
$h \in T^\omega$, namely the centralizer of $h$,
$\cZ(h)=\{ z\in G|\,zhz^{-1}=h \}$.
Then the D-brane $\cC_\omega(h)$ is the homogeneous space $G/\cZ(h)$.
For a generic point $h$ in $T^\omega$, its centralizer is $T^\omega$ itself
to yield the D-brane of maximal dimension \cite{FFFS}.
If $\cZ(h)$ becomes larger than $T^\omega$, we call $h$ a singular point
and the resulting D-brane has lower dimension.
We will develop a general recipe of the centralizer enhancement for
general compact Lie groups and corresponding D-branes as a
consequence. For simplicity, we restrict ourselves to regular conjugacy
classes without twist ($\Omega=1$, $\omega=1$).

Let $h$ be a point in the maximal torus, which can be given by
exponentiating an element $X$ in the Cartan subalgebra
$\mh\subset\mg$: $h\equiv e^{2\pi iX}$.
$X$ is parametrized by a weight vector $\vec{\psi}$ such that
\be
X=\vec{\psi}\cdot\vec{H}\quad\mbox{where}\quad
\vec{H}=(H_1,\cdots,H_r) \quad\mbox{with}\quad H_i \in \mh,
\label{x1}
\ee
where $r \equiv {\rm rank}\, \mg$.
A given Lie algebra $\mg$ of dimension $d$ and rank $r$ has the Cartan
decomposition $\mg =\mh\oplus(\bigoplus_\alpha \mg_\alpha)$:
\be
[\vec{H}, \vec{H}\,] = 0,\quad
\left.[\vec{H}, E_{\pm \alpha_i}]\right.
=\pm\vec{\alpha}_i E_{\pm \alpha_i},\quad
\left.[E_{\alpha_i}, E_{-\alpha_j}]\right.=\delta_{ij}\,
\vec{\alpha}_i \cdot \vec{H},
\label{Lie}
\ee
where $\alpha_i$'s ($i=1,\cdots,\frac{1}{2}(d-r)$) denote positive
roots and corresponding root vectors in weight space are given by
$\vec{\alpha}_i$'s. The first $r$ roots $\{\alpha_1,\cdots,\alpha_r\}$
denote simple roots.
In order to discuss both long and short roots on the same footing,
we introduce the  scale invariant generators of Lie algebra
$\mathfrak{su}(2)_{\alpha_i}$:
\be
H_{\alpha_i}\equiv\frac{\vec{\alpha}_i \cdot \vec{H}}
{|\hspace{1pt}\vec{\alpha}_i|^2},\quad
e_{\pm \alpha_i}\equiv\frac{E_{\pm
\alpha_i}}{|\hspace{1pt}\vec{\alpha}_i|}.
\label{norm}
\ee
These can be identified respectively with the $\rSU(2)$ spin operators,
$J^3$, $J^\pm$, regardless of whether the corresponding root $\alpha_i$
is long or short.

One first notice that $\exp\,(2\pi i n J_3)$ with arbitrary integers $n$
commute with all generators of the Lie algebra $\msu(2)$ and so is
$\exp\,(2\pi i nH_{\alpha_i})$ in the subgroup $\rSU(2)_{\alpha_i}$.
We first decompose $X=\vec{\psi}\cdot\vec{H}$ into $H_{\alpha_i}$ direction
and its orthogonal complement:
$X\equiv (\vec{\alpha}_i\cdot\vec{\psi}) H_{\alpha_i}+X^\perp$.
One can show that $[X^\perp, E_{\pm\alpha_i}]=0$ by using the Lie algebra
in Eq.\ (\ref{Lie}). Then $h\equiv e^{2\pi iX}$ can be factorized:
\be
h= e^{2\pi i X}=h^\perp
\exp\left[2\pi i(\vec{\alpha}_i\cdot\vec{\psi}) H_{\alpha_i}\right],
\ee
where $h^\perp \equiv e^{2\pi i X^\perp}$ commutes with $\msu(2)_{\alpha_i}$.
Hence $h$ commutes with $E_{\pm \alpha_i}$ if $\vec{\psi}$ is located on
any of the hyperplanes $\vec{\alpha}_i\cdot\vec{\psi}\in\Z$ which are
perpendicular to the root vector $\vec{\alpha}_i$.
On those hyperplanes, $\rU(1)_{\alpha_i}$ in the centralizer is enhanced
to $\rSU(2)_{\alpha_i}$ and $\cZ(h)$ becomes $\rSU(2)\otimes\rU(1)^{r-1}$.
Notice that the rank of the centralizer is preserved under these
enhancements.

Now we introduce the fundamental weight vectors
$\{\vec{\mu}_1,\cdots,\vec{\mu}_r\}$ as a basis of weight space.
They are defined by
\be
\frac{2\,\vec{\alpha}_i \cdot\vec{\mu}_j}
{|\hspace{1pt}\vec{\alpha}_i|^2}=\delta_{ij}, \label{fund}
\ee
where $\vec{\alpha}_i$'s are restricted to simple roots only.
Long root length is $\sqrt{2}$ as usual.
Under the decomposition
\be
\vec{\psi}=\sum_{i=1}^r \psi_i \,\vec{\mu}_i \label{Hm},
\ee
the coordinates $\psi_i$'s can be calculated to be
\be
\psi_i =l_i\,(\vec{\alpha}_i \cdot\vec{\psi})
\quad\mbox{with}\quad l_i = \frac{2}{|\hspace{1pt}\vec{\alpha}_i|^2}.
\label{psi}
\ee
The last formula can be used not only for simple roots ($i\le r$)
but also for all other positive roots ($i>r$).
Note that $l_i$ is given by a positive integer:
$l_i =1$ for long roots, whereas for short roots $l_i=2$ for $G=B_r$,
$C_r$, $F_4$ and $l_i=3$ for $G=G_2$.
The hyperplanes $\vec{\alpha}_i\cdot\vec{\psi}\in\Z$ with the symmetry group
$\rSU(2)_{\alpha_i}$ are specified by $\psi_i \in l_i\,\Z$.
For any non-simple root $\alpha_m$,  $\psi_m $ is given by a certain linear
combination of weight space coordinates $\{\psi_1,\cdots,\psi_r\}$
such that $\psi_m\in l_m\,\Z$ describe hyperplanes orthogonal to
the non-simple root vector $\vec{\alpha}_m$.

In terms of the coordinates $\psi_i$'s, the hyperplane with
$\rSU(2)_{\alpha_i}$ symmetry is written as
\be
P_{\alpha_i,\,n_i}\equiv\{\vec{\psi}\,|\,\psi_i =l_i\,n_i,\,n_i\in\Z\},
\ee
regardless of whether the root $\alpha_i$ is simple or non-simple.
The central lattice is a lattice in the weight space generated by
the $r$ vectors $l_i\,\vec{\mu}_i$ $(i=1,\cdots,r)$.
It is then obvious that the intersection points of the hyperplanes
for all the simple roots compose the central lattice.
Since the simple root system generates the whole of the group $G$,
all the points on the central lattice are mapped to the center of
the group under exponentiation, justifying the name of the lattice.
Consequently, the central lattice points correspond to D0-branes.

A mirror reflection on $P_{\alpha_i,\,n_i}$ is nothing but
the action of the extended Weyl group, the semi-direct product of
Weyl group and the translation on the co-root lattice.
Meanwhile, the fundamental domain of the extended Weyl group
(Weyl domain) is given by the minimal region surrounded by
all possible hyperplanes $P_{\alpha_i,\,n_i}$'s
($i=1,\cdots,\frac{1}{2}(d-r)$).
Consequently, a unit cell of the central lattice, $0\le\psi_i\le l_i$
$(i=1,\cdots,r)$, is further decomposed into Weyl domains.
As we will see later, not all the vertices of a Weyl
domain correspond to D0-branes.

The symmetry enhancement is therefore completely fixed by the position
of the D-brane in the weight space ($\vec{\psi}$).
For any simple Lie group $G$, the classification of the D-branes
can be described in terms of the Dynkin diagram as follows \cite{IS}.
Suppose the D-brane location $\vec{\psi}$ is belonging to the
intersection of $k$ hyperplanes.
Out of a Dynkin diagram of $G$ of rank $r$, we take away $r-k$ roots (circles)
corresponding to U(1)'s which are not enhanced to SU(2)'s.
Then the Dynkin diagram becomes the disjoint union of $\kappa=r-k+1$
sub-diagrams, each of which corresponds to a subgroup $G_{(i)}$
of the original group $G$.
Then the centralizer is given by $\prod_{i=1}^\kappa G_{(i)} \times
\rU(1)^{r-k}$. Finally the corresponding
D-brane $D(\vec{\psi})$ is given by the coset
\be
D(\vec{\psi})=G/\prod_{i=1}^{\kappa} G_{(i)} \times \rU(1)^{r-k},
\quad{\rm if}\quad
\vec{\psi} \in \bigcap_{i=1}^k P_{\alpha_i,n_i}.
\ee

Using this rule, one can write down the D-branes and their
dimensions explicitly.
For $B_r$-series, the centralizer is given by
\be
\cZ(h)=\rU(1)^{r-k} \times B_{r_1} \times \prod_{i=2}^{r-k+1}A_{r_i},
\label{enpatt}
\ee
allowing $B_0 =C_0 =\{0\}$, $B_1 =\rSO(3)$, $C_1 =\rSU(2)$.
The same formula holds for $C_r$-series if we replace $B_{r_1}$ with
$C_{r_1}$. The dimension of a D-brane in $B_r$-, $C_r$-series is given by
\be
p= 2\,(r^2-r_1^2)-\sum_{i=2}^{r-k+1}r_i\,(r_i+1),
\ee
where we have used $k=\sum_{i=1}^{r-k+1}r_i$.
Notice that it is manifestly an even integer,
a fact that can be related to the
existence of the symplectic structure  on the (co-)adjoint orbits.
Similarly the possible patterns of the centralizer in $F_4$ are
$F_4$, $\rU(1)\times C_3$, $\rU(1)\times B_3$,
$\rU(1)\times A_1 \times A_2$, $\rU(1)^2\times C_2$, $\rU(1)^2\times A_2$,
$\rU(1)^2\times A_1^2$, $\rU(1)^3\times A_1$, $\rU(1)^4$ where
$B_3$ must be identified as ${\rm Spin}(7)$ according to $\pi_1(F_4)=\{0\}$.

However, the minimal non-trivial block  $B_2$ (or $C_2$) arising
in the above enhancement patterns can be further decomposed into
its subgroups depending on the position $\vec{\psi}$ inside the
$B_2$ (or $C_2$) subspace. Therefore we need to work out the Weyl domains
in $B_2$ (or $C_2$) explicitly to complete the list of
centralizer enhancement.
In the next section, we study D-brane Zoo of all non-simply-laced
rank 2 groups, namely $B_2$, $C_2$, $G_2$.

\noindent
{\bf 3. D-brane Zoo}
\vspace*{3pt}

\begin{figure}
\begin{center}
\begin{minipage}[t]{5.5cm}
\centerline{\hbox{\psfig{file=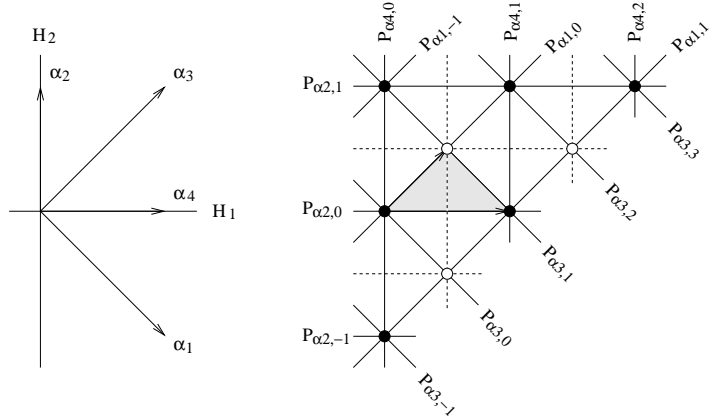,height=5.5cm}}}
\end{minipage}
\caption{Root vectors and a Weyl domain in SO(5) (Sp(2)).
Filled circles are central lattice points with SO(5) (Sp(2)) enhancement.
Every unfilled circle has the centralizer SO(4) ($\rSU(2)\times\rSU(2)$).}
\label{fig:so5w}
\end{center}
\end{figure}

{\it $B_2$ and $C_2$:}
Since the root systems for $B_2$ and $C_2$ are the same, we only have to
discuss $B_2$ as long as we classify the types of D-branes.
We will work on the fundamental representation of $B_2$.
Ten generators of $B_2=\rSO(5)$ are given by 5$\times$5 angular momentum
matrices:
$(M_{ab})_{cd} \equiv -i(\delta_{ac}\delta_{bd}-\delta_{ad}\delta_{bc})$.
One can choose Cartan generators as $(H_1,H_2)=(M_{12},M_{34})$.
The simple roots and fundamental weights of $\mathfrak{so}(5)$ are
\be
\begin{array}{ll}
\vec{\alpha}_1 = (1,-1),\quad &
\vec{\alpha}_2 = (0,~1),
\vspace{12pt}\\
\vec{\mu}_1 = (1,~0),\quad &
\vec{\mu}_2 = \left(\dfrac{1}{2},~\dfrac{1}{2}\right).
\end{array}
\ee
All other positive roots are
$\vec{\alpha}_3 = \vec{\alpha}_1 +2\vec{\alpha}_2$,
$\vec{\alpha}_4 = \vec{\alpha}_1 +\vec{\alpha}_2$.
We notice that $\vec{\alpha}_1$ and $\vec{\alpha}_3$ are long roots with
length $\sqrt{2}$, while $\vec{\alpha}_2$ and $\vec{\alpha}_4$ are short
roots with length 1 (See figure \ref{fig:so5w}).
For convenience, we diagonalize the Cartan generators $\vec{H}$ to yield
\be
H_1 ~=~{\rm diag}(0,1,0,-1,0), \quad H_2 ~=~{\rm diag}(1,0,0,0,-1).
 \label{cartan}
\ee

Recall that the weight vector $\vec{\psi}$ obeys the decomposition
in Eq.\ (\ref{Hm}). Any point $X \equiv \vec{\psi}\cdot\vec{H}$ on weight
space is specified by the coordinates $(\psi_1, \psi_2)$ in Eq.\
(\ref{psi}):
\be
X~=~{\rm diag}\!\left(\dfrac{\psi_2}{2},\psi_1+\dfrac{\psi_2}{2},0,
-(\psi_1+\dfrac{\psi_2}{2}),-\dfrac{\psi_2}{2}\right),
\label{so5X}
\ee
from which
one can see that if $\psi_1$ and $\psi_2 /2$ are integers,
$h \equiv e^{2\pi i X}$ becomes the identity matrix to enhance the centralizer
$\cZ(h)$ to SO(5). The points specified by both $\psi_1 \in \Z$
and $\psi_2 \in 2\Z$ are indeed the points on the central lattice of SO(5)
as discussed before.\footnote{
In fact, the central lattice of SO(5) is also the integral lattice.}
Note also that $l_1=1$ and $l_2=2$ in SO(5).

In order to determine Weyl domains in weight space, one has to know all
the possible enhancement lines given by the coordinates $\psi_i$'s.
By definition in Eq.\ (\ref{psi}), $\psi_3$ and $\psi_4$ are obtained as
\be
\psi_3 = \psi_1 +\psi_2,\quad \psi_4 = 2\psi_1 +\psi_2.
\ee
Since $\vec{\alpha}_3$ is a long root and $\vec{\alpha}_4$ is a short one,
the enhancement lines of $\rSU(2)_{\alpha_3}$ and $\rSU(2)_{\alpha_4}$ are
given by $\psi_1 +\psi_2 \in\Z$ and $2\psi_1 +\psi_2 \in 2\Z$ respectively.
Consequently, every Weyl domain is given by a triangle bounded by one long
edge of short root enhancement, either $\psi_2\in 2\Z$ or $\psi_4\in 2\Z$,
and two short edges of long root enhancement $\psi_1\in\Z$ and $\psi_3\in\Z$.
Each unit cell of the SO(5) central lattice is further decomposed into
four Weyl domains as shown in figure \ref{fig:so5w}.
Three lines of $\psi_1=0$, $\psi_2=0$, $\psi_3=1$ cut out a Weyl domain
of SO(5) weight space. An intersection point of two short edges $\psi_1=0$
and $\psi_3=1$ has SU(2)$\times$SU(2) symmetry corresponding to two
orthogonal long roots $\alpha_1$ and $\alpha_3$. Two end points of long
edge $\psi_2=0$ are located on the central lattice so that they have the
full symmetry SO(5).

This can be directly checked by using the matrix $X$ in Eq.\ (\ref{so5X}).
Suppose $\rSO(5)$ is acting on the space with coordinates
$(x_1,\cdots,x_5)$. Denote an $n\times n$ submatrix of a parent
$5\times 5$ matrix by $(i_1,\cdots,i_n)$ indicating the subspace
$(x_{i_1},\cdots,x_{i_n})$ on which the submatrix is acting.
As shown in appendix B.3 in \cite{IS}, $\rSU(2)_{\alpha_1}$ for a long root
is generated by two ${\bf 2}$'s of SU(2) specified by $(1,2)$, $(4,5)$ blocks,
whereas $\rSU(2)_{\alpha_2}$ for a short root is generated by a ${\bf 3}$ of
SU(2) given by $(1,3,5)$ block. Similarly, $\rSU(2)_{\alpha_3}$ is generated
by two ${\bf 2}$'s of $(1,4)$, $(2,5)$ blocks, while $\rSU(2)_{\alpha_4}$
is generated by a ${\bf 3}$ of $(2,3,4)$ block.
For the point $(\psi_1,\psi_2)=(1,1)$, $h\equiv e^{2\pi i X}$ is now
given by the diagonal matrix:
\be
h ~=~{\rm diag}(-1,-1,1,-1,-1)
\ee
which is proportional to the identity matrix both in $(1,2)$, $(4,5)$
blocks and in $(1,4)$, $(2,5)$ blocks, so that both $\rSU(2)_{\alpha_1}$
and $\rSU(2)_{\alpha_3}$ become symmetry of the $h$.

However, the same $h$ contains the identity matrix in $(1,2,4,5)$ block also.
Thus the centralizer $\cZ(h)$ seems to be SO(4) acting on $(1,2,4,5)$ block,
rather than $\rSU(2)\times\rSU(2)$. As we will see later,
the centralizer generated by two long roots
$\alpha_1$ and $\alpha_3$ depends on the global
topology of the parent group $G$, namely $\pi_1(G)=\{0\}$ or $\Z_2$.
For $G=\rSp(2)$, $\alpha_1$ and $\alpha_3$ generate
${\rm Spin}(4)\cong\rSU(2)\times\rSU(2)$, whereas for $G=\rSO(5)$
they generate SO(4).

\begin{figure}
\begin{center}
\begin{minipage}[t]{6cm}
\centerline{\hbox{\psfig{file=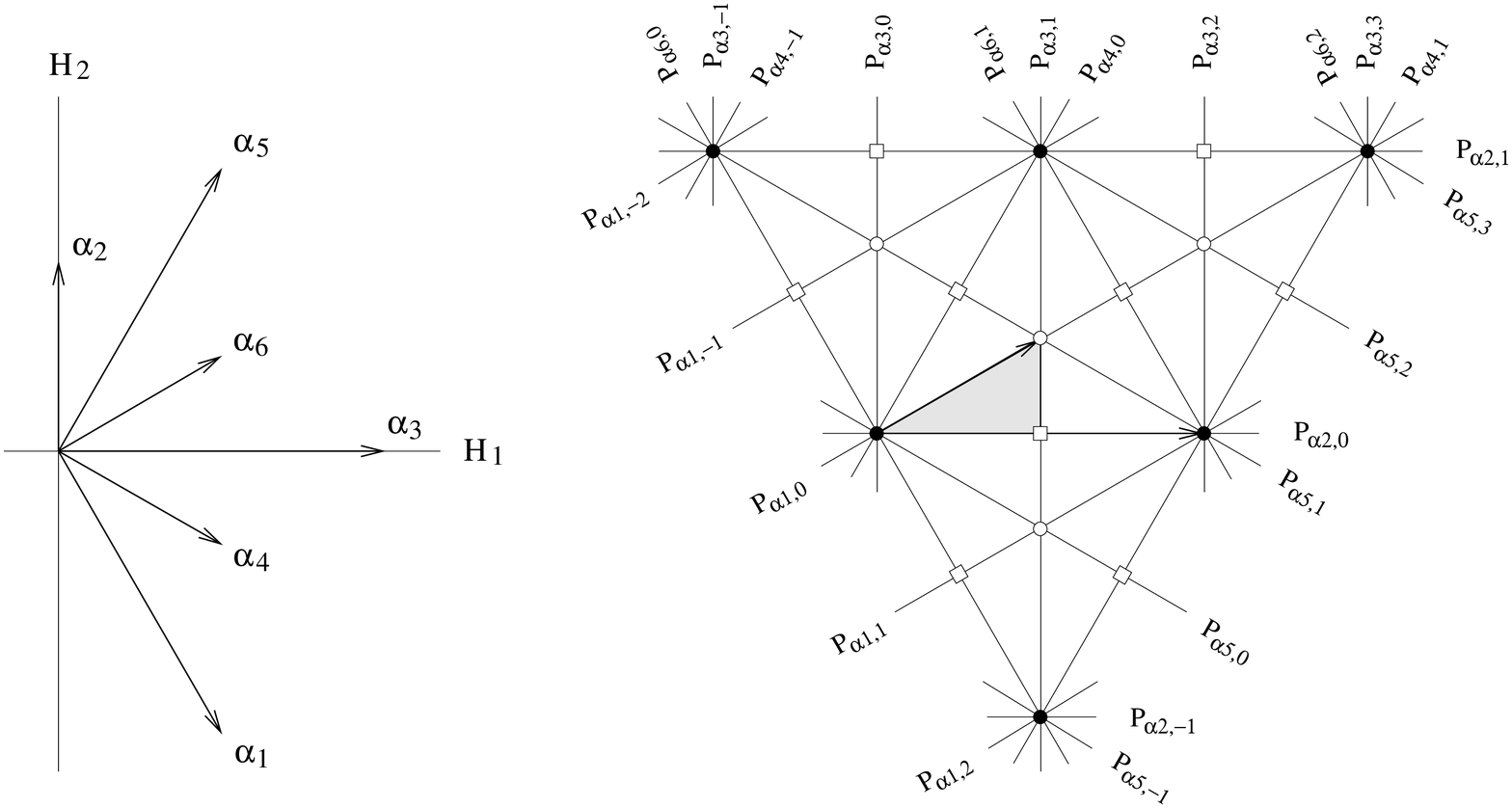,height=6cm}}}
\end{minipage}
\caption{Root vectors and a Weyl domain in G$_2$.
Filled circles are central lattice points with $G_2$ enhancement.
Every unfilled circle has the centralizer SU(3). The intersection
points indicated by squares have $\rSU(2)\times\rSU(2)$ enhancement.}
\label{fig:g2w}
\end{center}
\end{figure}

{\it $G_2$:}
The fundamental representation of $G_2$ is 7-dimensional and
${\bf 7}$ of $G_2$ goes to ${\bf 3 +\bar{3}+1}$ in an SU(3) subgroup.
As shown in \cite{GG}, the 14 generators of $G_2$ are
given by certain linear combinations of the 21 generators of $\rSO(7)$.
After appropriate diagonalization, the first 8 generators are given by
7$\times$7 matrices which are direct sum of
$\lambda_a \oplus\,[-\lambda_a]^T \oplus\,0$ where $\lambda_a$'s
are Gell-Mann matrices as expected from the branching rule
${\bf 7 \to 3 +\bar{3}+1}$ (See appendix B.5 in \cite{IS} also).
Recalling that SU(3) is the regular maximal subgroup of $G_2$, one can
choose Cartan generators of $G_2$ as the same as those of SU(3), namely
\be
H_1 =\frac{1}{\sqrt{2}}\left(\lambda_3 \oplus [-\lambda_3]^T \oplus 0\right),
\quad
H_2 =\frac{1}{\sqrt{2}}\left(\lambda_8 \oplus [-\lambda_8]^T \oplus 0\right).
\ee
The simple roots and fundamental weights of $\mg_2$ are
\bea
\begin{array}{ll}
\vec{\alpha}_1 = \left(\dfrac{1}{\sqrt{2}},-\sqrt{\dfrac{3}{2}}\,\right),
\quad &
\vec{\alpha}_2 = \left(0,~\sqrt{\dfrac{2}{3}}\,\right),
\vspace{12pt}\\
\vec{\mu}_1 = (\sqrt{2},~0),
\quad &
\vec{\mu}_2 = \left(\dfrac{1}{\sqrt{2}},~\dfrac{1}{\sqrt{6}}\right).
\end{array}
\eea
All other positive roots are
$\vec{\alpha}_3 = 2\vec{\alpha}_1 +3\vec{\alpha}_2$,
$\vec{\alpha}_4 = \vec{\alpha}_1 +\vec{\alpha}_2$,
$\vec{\alpha}_5 = \vec{\alpha}_1 +3\vec{\alpha}_2$,
$\vec{\alpha}_6 = \vec{\alpha}_1 +2\vec{\alpha}_2$.
We notice that $\vec{\alpha}_1$, $\vec{\alpha}_3$, $\vec{\alpha}_5$
are long roots with length $\sqrt{2}$, while $\vec{\alpha}_2$,
$\vec{\alpha}_4$, $\vec{\alpha}_6$ are short roots with length
$\sqrt{2/3}$ (See figure \ref{fig:g2w}).

For non-simple roots $\alpha_m$'s
($m=3,\cdots,6$),  $\psi_m$'s  are given by
\be
\psi_3=2\psi_1+\psi_2,\quad \psi_5=\psi_1+\psi_2,
\quad\psi_4=3\psi_1+\psi_2,\quad
\psi_6=3\psi_1+2\psi_2.
\ee
Notice that the long to short root ratio is 3 so that $l_1=1$, $l_2=3$.
The symmetry enhancement lines are given by
$\psi_l \in \Z$ for $l=1$, 3, 5 and $\psi_s \in 3\Z$ for $s=2$, 4, 6.
A unit cell of central lattice is the parallelogram spanned by
$\vec{\mu}_1$ and $3\vec{\mu}_2$ while the fundamental domain is surrounded
by $P_{\alpha_1,\,0}$, $P_{\alpha_2,\,0}$, $P_{\alpha_3,\,1}$.
The fundamental domain (or Weyl domain) is 1/12 of the unit cell
as shown in figure \ref{fig:g2w}.

All the long roots, $\alpha_1$, $\alpha_3$, $\alpha_5$, compose
an SU(3) subgroup of $G_2$ as shown in figure \ref{fig:g2w}.
Hence every intersection point where only three lines of long root
enhancement meet has a symmetry group SU(3)
as depicted by circles in figure \ref{fig:g2w}. The corresponding D-brane
becomes 6-dimensional one given by $G_2/\rSU(3)$.
We also recognize that all the short roots $\alpha_2$, $\alpha_4$, $\alpha_6$
compose another SU(3) subgroup of $G_2$. However, every intersection
point of short root enhancement lines is necessarily located on the central
lattice of $G_2$ (filled circles in figure \ref{fig:g2w}) and therefore
has the symmetry group $G_2$, not SU(3). Finally, we can choose three pairs
of long and short roots orthogonal to each other, namely
$(\alpha_1,\alpha_6)$, $(\alpha_3,\alpha_2)$, $(\alpha_5,\alpha_4)$.
These pairs correspond to the intersection points depicted by squares
in figure \ref{fig:g2w} where the symmetry group is enhanced to
$\rSU(2)\times\rSU(2)$.

\vspace*{12pt}
\noindent
{\bf 4. Classical moduli space and global issues}
\vspace*{3pt}

We have seen that symmetry enhancement is related to the central
lattice. Now  we introduce two other lattices, namely the integral
lattice and the co-root lattice, to define the moduli space of
D-branes and discuss the global issues of it.
\begin{itemize}
\item Integral lattice (IL) : It is defined as the inverse image of
the identity of $G$ under the exponentiation \cite{adams}.
The unit cell of the integral lattice can be identified with
the moduli space, since it is in one to one correspondence with
the maximal torus.
\item Co-root lattice (CL) : a sublattice of central lattice that is
generated by the co-roots. For a root $\alpha$, corresponding co-root
$\alpha^\vee$ is defined as $2\alpha/|\hspace{1pt}\alpha|^2$.
This lattice can also be defined as the image of the
origin under the reflections of the hyperplanes $P_{\alpha_i,n}$.
\end{itemize}

\begin{figure}
\begin{center}
\begin{minipage}[t]{4.5cm}
\centerline{\hbox{\psfig{file=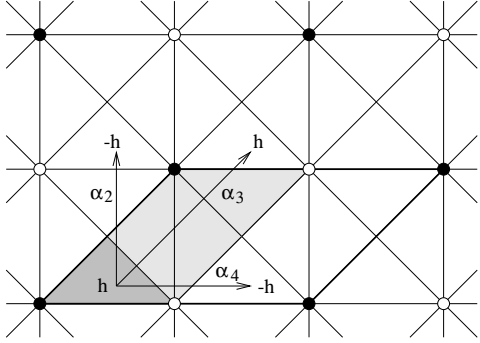,height=4.5cm}}}
\end{minipage}
\caption{The integral lattice of Sp(2) consists of all the filled
circles and coincides with the co-root lattice. The group manifold
Sp(2) is therefore simply-connected. In SO(5), both of filled and
unfilled circles compose the integral lattice, whereas the co-root
lattice is given by the filled circles only.}
\label{fig:global}
\end{center}
\end{figure}

{\it $B_2$ and $C_2$:}
The group space $B_2 =\rSO(5)$ is isomorphic to $C_2 =\rSp(2)$ locally
but not globally. In fact, $\rSO(5)\cong\rSp(2)/\Z_2$.
$B_2$ and $C_2$ share the same central and co-root lattices,
while they have different integral lattices. It turns out that
IL of $B_2$ is identical to the central lattice while that of
$A_2$ or $C_2$ is equal to the co-root lattice.
Now one can show that \cite{adams}
\be
\pi_1(G) = {\rm IL}/{\rm CL}. \label{pi1g}
\ee
Using Eq.\ (\ref{pi1g}), it is easy to show
that $\pi_1(\rSO(5))=\Z_2$, while $\pi_1(\rSp(2))=\{0\}$.
See figure \ref{fig:global}.
The two to one relation between the two groups can be understood from
the fact that the inverse image of the maximal torus, or a unit cell of IL,
of the Sp(2) is double size of the SO(5) when we draw them in the same
plane with the same normalization where the long root length is $\sqrt{2}$.

In fact two points $X$ and $X'$ related by translation with any short
root vector are mapped to the same point in SO(5) maximal torus but
different points in Sp(2) maximal torus. The difference is by $-1$ factor
since $\exp(2\pi i \vec{\alpha}_s\cdot\vec{H})=-I$ for any short root
$\alpha_s$ ($I$ is the identity element in Sp(2)).
This is because short root vectors are elements of the integral lattice
in SO(5) but not in Sp(2). Note that the central lattice is decomposed
into the co-root lattice (CL) and its complement lattice (CL$^\prime$).
Both CL and CL$^\prime$ compose IL in SO(5), whereas IL of Sp(2) is
given by CL only. In Sp(2), CL and CL$^\prime$ are exponentiated to be
different centers $I$ and $-I$, respectively. See figure \ref{fig:global}.
Thus, the $\Z_2$ discussed so far can be identified with the center of Sp(2)
to get the group space SO(5) as an orbifold $\rSp(2)/\Z_2$.
Things are the same for the subgroup generated by $\alpha_1$ and
$\alpha_3$. They generate SO(4) if the parent group is SO(5),
whereas they generate ${\rm Spin}(4)$ as a subgroup of Sp(2).

One can also see that the $\Z_2$ symmetry acts on the weight space as
the mirror reflections on the lines $\psi_s = 1$ mod 2 ($s=2,4$),
indicated by dashed lines in figure \ref{fig:so5w}.
All the CL points satisfy $\psi_2 \pm \psi_4 = 0$ mod 4, while all the
CL$^\prime$ points satisfy $\psi_2 \pm \psi_4 = 2$ mod 4.
The mirror reflections
on the lines $\psi_s = 1$ mod 2 ($s=2,4$) are given by
$\psi_s \to 2 - \psi_s$ ($s=2,4$), respectively. Under the reflections
CL is mapped to CL$^\prime$ and vice versa.
The SO(5) weight space is therefore invariant under the $\Z_2$ reflections.
Two D-branes related by the $\Z_2$ reflection are different in Sp(2)
but the same in SO(5). Thus we can expect that in SO(5) there exist
$\Z_2$ invariant D-branes located on the reflection lines and
such a D-brane arises as an unoriented one
$\rSO(5)/(\rSO(2)\!\times\!\rO(2))$. This can be seen by looking at
$h=e^{2\pi iX}$ with $X$ in Eq.\ (\ref{so5X}), which gives
$h={\rm diag}(z,-1,+1,-1,\bar{z})$ for the lines $\psi_4=1$ mod 2.
Each of them has an $\rO(2)$ rather than $\rSO(2)$ enhancement in (2,4) block.
Similarly, any of the lines $\psi_2=1$ mod 2 has an extra $\Z_2$
enhancement in $(1,5)$ block.

{\it $G_2$:}
By looking at the explicit form of 7$\times$7 matrix
$X=(\psi_1 \vec{\mu}_1 +\psi_2 \vec{\mu}_2)\cdot\vec{H}$, we notice that
the integral lattice of $G_2$ is at the same time the central lattice
depicted by filled circles in figure \ref{fig:g2w} where heperplanes of
all positive roots intersect. Note also that the central lattice
coincides with the co-root lattice in $G_2$. Eq.\ (\ref{pi1g}) therefore
concludes that $\pi_1(G_2)=\{0\}$ and the group manifold $G_2$ is
simply-connected. If $G_2$ is restricted to its SU(3) subspace generated
by long roots, one can think of both filled and unfilled circles in
figure \ref{fig:g2w} as the central lattice points of SU(3).
Then the integral lattice of $G_2$ is also that of the SU(3) and coincides
with the SU(3) co-root lattice to conclude that $\pi_1(\rSU(3))=\{0\}$.

\vspace*{12pt}
\noindent
{\bf 5. Quantum moduli space and global group structures}
\vspace*{3pt}

Quantum mechanically, the single-valuedness of the path integral
of the level $k$ boundary WZW model gives two quantization conditions:
one from the presence of $H$-monopoles, the other from that of
$F$-monopoles. The former condition gives the quantization of level $k$,
while the latter gives the condition that
$k\,\vec{\psi}\cdot\vec{\alpha}_i^{\,\vee}$
should be an integer $n_i$ for all positive roots $\alpha_i$, or equivalently
$k\,\vec{\psi}$ must be a highest weight \cite{FFFS,G,S2}.
The presence of $H$-monopoles also gives the condition that $k\,\vec{\psi}$
should be defined only modulo $k$.

By using Eq.\ (\ref{fund}) saying that the co-roots are dual to the
fundamental weights, the stable position of a D-brane is given by
\be
 k\,\vec{\psi}=\sum_{i=1}^r n_i \,\vec{\mu}_i, \label{qm}
\ee
with the coefficients $(n_1,\cdots,n_r)$ for simple roots.
We call the set of those points satisfying Eq.\ (\ref{qm}) the quantum
moduli space. As we will show shortly, the co-roots generate the homology
group $H_2(G/T;\Z)$, namely
\be
H_2(G/T;\Z) \cong {\rm CL},\label{qmhomo}
\ee
where CL is the co-root lattice of $G$. Therefore Eq.\ (\ref{qm})
says that the highest weight $k\,\vec{\psi}$ must be an element of
the cohomology group $H^2(G/T;\Z)$ dual to $H_2(G/T;\Z)$.

Now let us prove Eq.\ (\ref{qmhomo}).
We start from the exact homotopy sequence \cite{NS}
\be
\cdots \to \pi_2 (G) \to \pi_2 (G/T) \to \pi_1 (T) \to \pi_1 (G)\to
\pi_1 (G/T) \to \pi_0 (T) \to \cdots. \label{homoseq}
\ee
From $\pi_0(T)=\{0\}$ and $\pi_1(G)=\{0\}$ or $\Z_2$,
the sequence implies that $\pi_1(G/T)$ is either $\{0\}$ or $\Z_2$.
It can not be ${\bf Z}_2$, since it would mean that
$\pi_{1} (T)=\{0\}$, which is not true.  Therefore we must have
$\pi_1(G/T)=\{0\}$. Then  $\pi_1(T)={\bf Z}^r$ is classified by
$\pi_1(G)={\bf Z}_2$ due to the surjectiveness of
$\Phi:\pi_{1} (T )\mapsto\pi_{1}(G)$.
From $\pi_2(G)=\{0\}$ for any compact connected Lie group $G$, $\pi_2 (G/T)$
can be identified with its image in $\pi_1(T)$.
Due to the exactness of the sequence, the image of $\pi_2 (G/T)$
is equal to the kernel of $\Phi$.
Thus we arrive at the non-trivial relation $\pi_2(G/T)={\rm Ker}\,\Phi$.
Since the first non-trivial homology group and the
first non-trivial homotopy group are isomorphic \cite{NS}, we get
$H_2 (G/T;\Z)={\rm Ker}\,\Phi$.
Every unit cell of the integral lattice of $G$ is exponentiated
to be the same maximal torus $T \subset G$ and it is obvious
$\pi_1(T)\cong{\rm IL}$. Recall $\pi_1(G)\cong{\rm IL/CL}$
as shown in Eq.\ (\ref{pi1g}).
Then the surjective mapping $\Phi$ can be rewritten as
\be
\Phi: {\rm IL}\mapsto{\rm IL/CL}
\ee
which immediately says ${\rm Ker}\,\Phi\cong{\rm CL}$ and
we arrive at Eq.\ (\ref{qmhomo}).

Although the condition (\ref{qm}) arises in the same way for SO(5) and Sp(2),
its meaning is different between two cases.
The condition says that half the Weyl domain enclosed by three lines
$\psi_1=0$, $\psi_2=0$, $\psi_4=1$ is discretized into a set of stable
points. This can be seen by setting $n_4=k$ in $k\,\psi_4=n_4$.
Hence the D-branes located on the $\Z_2$ reflection lines
$\psi_s = 1$ mod 2 ($s=2,4$) can be stable quantum mechanically.
Such a D-brane arises as an unoriented one $\rSO(5)/(\rSO(2)\times\rO(2))$
in SO(5), while as a generic one $\rSp(2)/\rU(1)^2$ in Sp(2).
Things are the same when we compare SO($N$)'s with their covering groups
Spin($N$)'s ($N\ge3$). For SO(3) as $B_1$ generated by a single short root,
an unoriented D-brane $\rSO(3)/\rO(2)$ can be stable as shown in \cite{S3}.

So far our discussion on quantum moduli space is just based on the
(co)homological consideration. According to the CFT analysis,
the condition (\ref{qm}) must be corrected in two ways \cite{FFFS}.
First, the level $k$ must be shifted to $k+x$, where $x$ is the dual
Coxeter number of the group $G$ and is given by $N-2$ for SO($N$),
$N+1$ for Sp($N$), 9 for $F_4$, 4 for $G_2$.
Second, the highest weight $k\,\vec{\psi}$ must be shifted by
the Weyl vector $\vec{\rho}$, that is defined as half the sum of all
positive roots. Thus the condition (\ref{qm}) is corrected to be \cite{FFFS}
\be
 (k+x)\,\vec{\psi} = \sum_{i=1}^r n_i\,\vec{\mu}_i +\vec{\rho}.
\label{exqm}
\ee
However, we should notice that the Weyl vector is always equal to the sum of
all fundamental weights, that is $\vec{\rho}=\sum_{i=1}^r \vec{\mu}_i$,
for any compact Lie group. Therefore the exact condition (\ref{exqm})
is not so different from the semi-classical one (\ref{qm}) except for
shifting $k\to k+x$ and $n_i \to n_i+1$. The singular D-branes are still
allowed quantum mechanically.

\vspace*{12pt}
\noindent
{\bf 6. Conclusions}
\vspace*{3pt}

In this paper, we gave a general prescription for the D-brane
classification according to the D-brane position in the
fundamental domain of the weight space. Utilizing the method of
iterative deletion in Dynkin diagram, we constructed all the
D-branes in compact non-simply-laced Lie groups. In $G_2$ case, we
found the D6-brane $G_2/\rSU(3)$, that is the 6-dimensional sphere
embedded in $\R^7$ spanned by imaginary octonions \cite{GW}. We
also described the global issues involved in the SO groups using
the integral and co-root lattices. The group space SO(5) can be
understood as an orbifold $\rSp(2)/\Z_2$ where $\Z_2$ is the
center of Sp(2). The $\Z_2$ invariant D-brane in SO(5) arises as
an unoriented D-brane $\rSO(5)/(\rSO(2)\!\times\!\rO(2))$. The
semi-classical condition (\ref{qm}) for the quantum moduli space
is not affected so much by the CFT corrections and the singular
D-branes can be stable even for the finite level $k$. However,
there is also another effect working at finite level: the brane
world volume is not sharply localized and becomes `fuzzy' as shown
in \cite{FFFS}. It might be interesting to see how the methods of
present paper can be used to discuss this effect or related
quantum symmetries discussed  in \cite{PS}. For future work, we
may extend our analysis in this paper to the classification of
twisted D-branes \cite{FFFS,S3} and also to D-branes in coset
spaces.

\vspace*{12pt}
\noindent
{\bf Acknowledgements:}
This work is supported in part by KOSEF 1999-2-112-003-5. It is also
supported in part by Hanyang University, Korea made in the program
year of 2001. The authors are grateful to J.\ Fuchs for his comments
on the earlier version of this paper.




\begin{thebibliography}{99}
\bibitem{AS} A.Yu.\ Alekseev and V.\ Schomerus,
Phys.\ Rev.\ D {\bf 60} (1999) 061901, {\tt hep-th/9812193}.
\bibitem{FFFS} G.\ Felder, J.\ Fr\"ohlich, J.\ Fuchs and C.\ Schweigert,
J.\ Geom.\ Phys.\ {\bf 34} (2000) 162, {\tt hep-th/9909030}.
\bibitem{S} S.\ Stanciu,
JHEP {\bf 0001} (2000) 025, {\tt hep-th/9909163}.
\bibitem{S3} S.\ Stanciu,
``An illustrated guide to D-branes in SU(3),''
{\tt hep-th/0111221}.
\bibitem{IS} T.\ Itoh and S.-J.\ Sin,
``A note on singular D-branes in group manifolds,''
{\tt hep-th/0206238}.
\bibitem{ishibashi} N.\ Ishibashi,
Mod.\ Phys.\ Lett.\ A {\bf 4} (1989) 251.
\bibitem{KO} M.\ Kato and T.\ Okada,
Nucl.\ Phys.\ B {\bf 499} (1997) 583, {\tt hep-th/9612148}.
\bibitem{GG} M.\ G\"{u}naydin and F.\ G\"{u}rsey,
J.\ Math.\ Phys.\ {\bf 14} (1973) 1651.
\bibitem{adams} J.F.\ Adams,
``Lectures on Lie Groups,'' (The University of Chicago Press, Chicago, 1969).
\bibitem{G} K.\ Gaw\c{e}dzki,
``Conformal field theory: a case study,''
{\tt hep-th/9904145}.
\bibitem{S2} S.\ Stanciu,
JHEP {\bf 0010} (2000) 015, {\tt hep-th/0006145}.
\bibitem{NS} C.\ Nash and S.\ Sen,
``Topology and Geometry for Physicists,'' (Academic Press, London 1983).
\bibitem{GW} M.\ G\"{u}naydin and N.P.\ Warner,
Nucl.\ Phys.\ B {\bf 248} (1984) 685.
\bibitem{PS}  J.\ Pawelczyk, H.\ Steinacker,
Nucl.\ Phys.\ B {\bf 638} (2002) 433, {\tt hep-th/0203110}.
\end{thebibliography}
\end{document}